\documentclass[sigconf]{acmart}
\copyrightyear{2020}
\acmYear{2020}
\setcopyright{acmcopyright}\acmConference[MM '20]{Proceedings of the 28th ACM
International Conference on Multimedia}{October 12--16, 2020}{Seattle, WA, USA}
\acmBooktitle{Proceedings of the 28th ACM International Conference on Multimedia
(MM '20), October 12--16, 2020, Seattle, WA, USA}
\acmPrice{15.00}
\acmDOI{10.1145/3394171.3414032}
\acmISBN{978-1-4503-7988-5/20/10}
\usepackage{xcolor,colortbl}
\usepackage{booktabs}
\usepackage{multirow}
\usepackage{rotating}
\usepackage{diagbox}
\usepackage{array}
\usepackage{enumerate}
\usepackage{amsfonts}
\usepackage{amsmath}
\usepackage{bookmark}
\usepackage{enumitem}
\usepackage{caption}

\usepackage[compact]{titlesec}
\begin{document}
\title{PiRhDy: Learning Pitch-, Rhythm-, and Dynamics-aware Embeddings for Symbolic Music}
\author{Hongru Liang}
\affiliation{
  \institution{Institute of Big Data}
  \institution{College of Computer Science}
  \institution{Nankai University}
}
\email{lianghr@mail.nankai.edu.cn}
\author{Wenqiang Lei}
\authornote{Corresponding author}
\affiliation{
  \institution{National University of Singapore}
}
\email{wenqianglei@gmail.com}

\author{Paul Yaozhu Chan}
\affiliation{
  \institution{Institute for Infocomm Research, A*STAR, Singapore}
}
\email{ychan@i2r.a-star.edu.sg}
\author{Zhenglu Yang}
\affiliation{
  \institution{Institute of Big Data}
  \institution{College of Computer Science}
  \institution{Nankai University}
}
\email{yangzl@nankai.edu.cn}
\author{Maosong Sun}
\affiliation{
  \institution{Tsinghua University}
}
\email{sms@tsinghua.edu.cn}
\author{Tat-Seng Chua}
\affiliation{
  \institution{National University of Singapore}
}
\email{chuats@comp.nus.edu.sg}

\begin{abstract}
Definitive embeddings remain a fundamental challenge of 
 computational musicology for symbolic music in deep learning today. Analogous to natural language,
 music can be modeled as a sequence of tokens. This motivates the majority of existing solutions to explore the utilization of word embedding models
 to build music embeddings. However, music differs from natural languages in two key aspects: (1) musical token is multi-faceted -- it comprises of pitch, rhythm and dynamics information;
 and (2) musical context is two-dimensional -- each musical token is dependent on
 both melodic and harmonic contexts.  
In this work, we
provide a comprehensive solution by proposing a novel framework named PiRhDy that integrates \underline{pi}tch, \underline{rh}ythm, and \underline{dy}namics information seamlessly. 
PiRhDy adopts a hierarchical strategy which can be decomposed into two steps: (1) token~(i.e., note event) modeling, which separately represents pitch, rhythm, and dynamics and integrates them into a single token embedding; and (2) context modeling, which utilizes melodic and harmonic knowledge to train the token embedding. 
A thorough study was made on each component and sub-strategy of PiRhDy.
We further validate our embeddings in three downstream tasks -- melody completion, accompaniment suggestion, and genre classification. Results indicate a significant advancement of the neural approach towards symbolic music as well as PiRhDy's potential as a pretrained tool for a broad range of symbolic music applications.

\end{abstract}



\maketitle
\section{Introduction}
Recent years have seen tremendous success in pretrained word embeddings. BERT~\cite{devlin2019bert} and GPT~\cite{radford2019language,brown2020language}), for example, have both brought great advancements to the progress of natural language processing~(NLP)~\cite{lei2020interactive,lei2020estimation,lei2018sequicity}.
Symbolic music processing is another domain which addresses other real-world applications such as
music generation~\cite{dong2018musegan} and music recommendation~\cite{damak2019seer}.
Just like NLP, the mining of meaningful information is equally important in this domain. 
In this work, we wish to design an effective framework to accurately represent the nature of symbolic music, and, in doing so, be able to embed key music information into a shared low-dimension space. 
In this way, the understanding of complex music can be formulated as a computational process of these representations.

Symbolic music may be thought of as an intermediary between notated music and musical sounds~\cite{kia2007inter}. 
It is similar to natural language in many aspects~\cite{mcmullen2004music,jackendoff2009parallels}. For example, both contain sequential tokens and are context-dependent. Hence, several recent efforts of symbolic music embeddings focus on investigating the potential application of word embeddings techniques~(e.g., CBOW and skip-gram~\cite{mikolov2013distributed, mikolov2013efficient}) 
towards
music. The surface form~(i.e., musical token) towards the concept of ``word'' in symbolic music involves fixed-length
slice (i.e., note event)~\cite{herremans2017modeling,chuan2020context}, a set of notes~(i.e., chord)~\cite{huang2016chordripple,madjiheurem2016chord2vec}, and a sequence of notes~(i.e., motif)~\cite{hirai2019melody2vec,alvarez2019distributed}. Amongst these, the training paradigm predicts either the center musical token from context~(CBOW) or the context from the center token~(skip-gram).

Although these works have produced seemingly promising music embeddings, they are far from satisfactory due to the inability to capture the special characteristics of music. In particular, music differs in two key aspects from natural languages: (1) musical tokens is a combination of multi-faceted features including pitch, rhythm, and dynamics; and (2) music is multi-dimensional constitutionally with its melodic context progressed in the horizontal axis and harmonic context organized in the vertical axis. Thus, we argue that two fundamental problems need to be solved: 1)~\textit{how to leverage pitch, rhythm, and dynamics 
information simultaneously}; and 2) \textit{how to encode both melodic and harmonic contexts comprehensively}.

\begin{table}[t]
    \centering
    \caption{Existing embeddings w.r.t. musical information~(the pitch, rhythm and dynamics columns) and context~(the melody and harmony columns) they model.}
    \scalebox{0.9}{
        \begin{tabular}{|l|r|r|r||l|r|}
            \toprule
            \multicolumn{1}{|l|}{music embeddings} & \multicolumn{1}{c|}{\begin{sideways}pitch\end{sideways}} & \multicolumn{1}{c|}{\begin{sideways}rhythm\end{sideways}}  & \multicolumn{1}{c||}{\begin{sideways}dynamics\end{sideways}} & \multicolumn{1}{c|}{\begin{sideways}melody\end{sideways}} & \multicolumn{1}{c|}{\begin{sideways}harmony\end{sideways}} \\
            \midrule
            chordripple~\cite{huang2016chordripple} &  \checkmark       &       &       & \checkmark &  \\
            \midrule
            chord2vec~\cite{madjiheurem2016chord2vec} &{\checkmark}   &             &       & \checkmark &  \\
            \midrule
            \citeauthor{herremans2017modeling}~\cite{herremans2017modeling} & \checkmark             &  &       & \checkmark & \multicolumn{1}{l|}{\checkmark} \\
            \midrule
            \citeauthor{chuan2020context}~\cite{chuan2020context} &\checkmark             &  &       & \checkmark & \multicolumn{1}{l|}{\checkmark} \\
            \midrule
            melody2vec~\cite{hirai2019melody2vec} & \multicolumn{1}{l|}{\checkmark} & \multicolumn{1}{l|}{\checkmark} &             & \checkmark &  \\
            \midrule
            \citeauthor{alvarez2019distributed}~\cite{alvarez2019distributed} & \checkmark           &  &       & \checkmark &  \\
            \bottomrule
            \end{tabular}%
        
      }
    \label{tab:related}%
  \end{table}%

To the best of our knowledge, there is still no unified framework that addresses these problems comprehensively. A brief summary of current music embeddings
are listed in Table~\ref{tab:related}. For information utilization, with the exception of melody2vec~\cite{hirai2019melody2vec}, existing methods focus on pitch information. Few consider the dynamics, which carries the variations in the loudness of music and is one of the most expressive elements of music~\cite{oore2018time}. This makes these methods unable to distill enough features for general tasks. Another observation is that aside from the note-event based approaches~\cite{herremans2017modeling,chuan2020context}, most embeddings do not model harmonic context, which contains the vertical knowledge of music. This leaves the vertical dimension of music unaccounted for, rendering them incapable of learning the complete knowledge from musical contexts.\par

Considering the limitations of existing solutions, we believe that it is critical to develop a framework that not only integrates multi-faceted features of musical tokens but also transfers knowledge from both melody and harmony into embeddings. Therefore, we propose a hierarchical framework consisting of two-stage modeling aligned with the fundamental problems. First, we design a token~(note event) modeling network to fuse pitch, rhythm, and dynamics features seamlessly. This network is built on our delicately designed musical vocabulary and consists of several efficient strategies to extract vital information. For example, pitch modeling is developed to fuse chroma and octave features into pitch information. Secondly, we build a context modeling network that can predict the probability distribution of music thoroughly. In this network, music embeddings are pretrained at the token-level~(
i.e., note event-level) context and fine-tuned at the sequence-level~(i.e., period-level and track-level) context. In this way, both short-term and long-term relations are encoded into the embeddings.

 Our study fundamentally contributes to neural symbolic music processing by providing pretrained embeddings. 
 The embeddings provide a bridge between arts and engineering ---  they are grounded with professional music theories and intuitions, and can serve as a plug-and-play tool for any downstream tasks as long as symbolic music modeling are required. To summarize, our contributions are:
\begin{itemize}[leftmargin=*]
    \item The first embeddings~(PiRhDy) that integrate \underline{pi}tch-, \underline{rh}ythm- and \underline{dy}namics-aware embeddings for symbolic music from both melodic and harmonic contexts.
    \item An extensive study of
    PiRhDy and demonstration of the necessity of integrating key features, the effectiveness of utilizing comprehensive contexts, and the robustness of our embeddings. 
    \item A thorough evaluation of the ability of PiRhDy embeddings to capture musical knowledge on tasks at different levels, that is, the sequence-level~(melody completion, accompaniment assignment) and song-level~(genre classification) tasks. 
\end{itemize}

\par

\section{Related work}
The low-dimensional embeddings in symbolic music can be separated into approaches based on
chord, note event and motif, which
correspond to the concept of ``word'' applied to music.

\textbf{Chord-based approaches}~\cite{madjiheurem2016chord2vec,huang2016chordripple} aim to learn chord~(a set of simultaneous notes) representations in the word2vec models. However, chords, from the accompaniment track, are purely aiding components of notes from the melody track. In other words, they only contribute to harmony. Hence, these works cannot generate universal embeddings for symbolic music. Besides, the study of chords requires all the chord attributes to be annotated and thus needs the help of experts.

\textbf{Note event-based approaches} treat music as an organized sequence of note events, which are the smallest unit of naturalistic music~\cite{herremans2017modeling,chuan2020context}. However, these works only train embeddings on the most frequent ``words'' to overcome the long-tail issues caused by huge vocabularies. For example, only 500 out of 4075 notes events are considered in~\cite{chuan2020context}, leading to incomplete learning of prior distribution and knowledge from a corpus.

\textbf{Motif-based approaches}~\cite{alvarez2019distributed,hirai2019melody2vec} keep tracks of the sequences of notes that may be referred to as motifs. Although motifs are the most similar in concept to words in natural language, there are no established dictionaries for motifs in music. Towards the study of motifs, \cite{alvarez2019distributed} redefines motif as fixed-length pitch intervals, which only cover melodic information. Alternatively, \cite{hirai2019melody2vec} extracts motifs from melody tracks using the generative theory of tonal music~\cite{lerdahl1996generative}, which is a rule-based approach. Either way, the learning procedure is limited in the melody track neglecting rich information in the accompaniment tracks. Moreover, similar to note-event based approaches, the long-tail distribution of motifs remains a problem. For example, more than half of the motifs cannot be sufficiently trained in melody2vec~\cite{hirai2019melody2vec}. \par
Besides contextual embeddings, 
another trend represents symbolic music as a sparse
matrix. Conventional \textbf{matrix-based approaches} transform MIDI files into pianorolls ~\cite{raffel2014intuitive, dong2018pypianoroll} (i.e. binary-valued time-pitch matrices). Specifically, if a bar is sliced into 96 time steps horizontally and 128 pitch values vertically, the size of pianoroll matrix of this bar is $96\times 128$~\cite{serra2012unsupervised}.
Notably, instead of manipulating data sequentially, \cite{chuan2018modeling} integrates knowledge from music theory~(i.e., tonnetz network) by formatting a slice of music into a binary-valued $4\times24\times12$ matrix. Thus, such representation naturally contains tonal relationships~(e.g., ordered according to the circle-of-fifths), yet at the cost of losing the temporal information among the notes in the slice. Both pianoroll and tonnetz representations are more likely to meet the challenge of limited computational and storage resources than low-dimensional dense embeddings.\par
 
\begin{table*}[t]
    \centering
    \caption{Overview of useful terms}
    \label{tab:term}
    \scalebox{0.8}{
    \begin{tabular}{|m{0.16\textwidth}|c|m{0.9\textwidth}|}
        \hline
        \textbf{term} & \textbf{notation} & \textbf{definition} \\
\hline
        note event &$n$& the playing of a note at a certain time span~(duration).\\
        \hline
        pitch& $P$ & a set of human-defined numbers describing the frequency degree of music sound, e.g., A4.\\
        \hline
        chroma & $c$ & a.k.a, pitch class, the octave-invariant value of a pitch. For example, both C3 and C4 refer to chroma C~($do$), and the pitch interval from C3 to C4 indicates an octave~(notated as $o$).\\
        \hline
        chord & $-$ & a set of two or more simultaneous notes. \\
        \hline
        duration & $-$ & the temporal length of a musical note, e.g., $1/4$.\\
        \hline
        onset & $-$ & the beginning of a musical note.\\
        \hline
        note state & $s$ & the current state of a note event, involving \textit{on}, \textit{hold}, and \textit{off} indicating the beginning, playing, and ending of a note, respectively.\\
        \hline
        inter-onset-interval & $i$ &  abbreviated to IOI, the duration between the onsets of two consecutive notes.\\
        \hline
        rhythm & $R$ &the temporal pattern of notes, related to note duration, note onset, and IOI.\\
        \hline
        dynamics &$D$ & the variation in loudness between notes or phrases.\\
        \hline
        velocity & $v$ & the dynamics markings~(e.g., ``$mf$'' and ``$p$'').\\
        \hline
        motif &$-$& a.k.a, motivate, a group of note events forming the smallest identifiable musical idea, normally one-bar long.\\
        \hline
        phrase &$ph.$& a group of motifs forming a complete musical sense, being normally four-bar long.\\
        \hline
        period &$per.$& a pair of consecutive phrases.\\
        \hline
        melody &$-$& a sequence of single notes from the melody track\footnote{In this paper, we assume there must be one and only one melody track of a song.}, which plays the most impressive sounds among the entire song. \\
        \hline
        harmony &$-$&the simultaneous or overlapping notes on accompaniment tracks to produce an overall effect along with the melody.\\

        \hline
    \end{tabular}
    }
    \label{tab:term}
\end{table*}
\begin{figure}[t]
    \centering
    \includegraphics[width=0.9\linewidth]{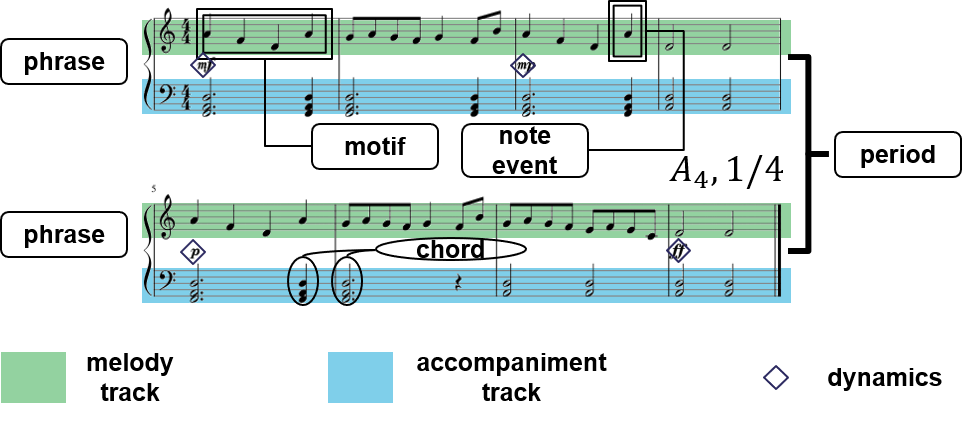}
    \caption{Example of music scores of a song, which consists of one period, a melody track and an accompaniment track.}
    \label{fig:structure}
\end{figure}

In addition to Table~\ref{tab:related}, we review the utilization of various information in matrix-based approaches. Both the pianoroll~\cite{raffel2014intuitive, dong2018pypianoroll} and tonnetz \cite{chuan2018modeling} can model melody and harmony.
However, the tonnetz only leverages pitch information. While the pianoroll utilizes both pitch and rhythm information apart from melody2vec~\cite{hirai2019melody2vec}. None of these methods untilize adequate information to encode music.

Altogether, there are four issues in existing approaches: cost of expert annotation, cost of computational and storage resources, incomplete prior distribution learning, and inadequate information learning. As a result, it remains challenging to 1)~\textit{leverage pitch, rhythm, and dynamics information simultaneously} and 2) \textit{encode both melodic and harmonic contexts comprehensively}. In contrast to the above-mentioned embeddings, we plan to learn distributed embeddings for symbolic music from scratch. We carefully construct a very concise vocabulary based on pitch, rhythm, and dynamics without regards to expert knowledge in order to learn robust embeddings from a limited corpus. We optimize the embeddings on both melodic and harmonic contexts in order to transfer both horizontal and vertical knowledge from music into the embeddings.

\section{Preliminary}
\label{sec:term}
    Table~\ref{tab:term} presents an overview of useful terms in this paper. Besides, we illustrate part of these terms in Figure~\ref{fig:structure}. In favour of~\cite{herremans2017modeling,chuan2020context}, we study music as note events
    but from a more delicate perspective~(see section~\ref{sec:method}). In practice, we need to integrate the pitch, rhythm, and dynamics information of note events. 
    
    First of all, pitch information is translates to chroma and the octave. For example, $A4$ can be expressed as the chroma $A$ on the $4$ octave group. As such, we define the pitch information as $P = \mathcal{F}_{P}(c, o)$, where $\mathcal{F}_{P}$ is the fusing function. 
    Second, the rhythm information is related to note duration, inter-onset-interval~(IOI), and note state. We only use IOI and note state values when modeling rhythm, as the note duration values 
    are
    derived completely from note events. Thus, we define the rhythm information as $R = \mathcal{F}_{R}(i, s)$, where $\mathcal{F}_{R}$ is the fusing function.
    Third, the dynamics information is expressed as velocity values. We formulate it as $D = \mathcal{F}_{D}(v)$, where $\mathcal{F}_{D}$ is the processing function. 
    
    Therefore, the study of note events is summarised in the modeling of chroma, octave, IOI, note state, and velocity features. This may be expressed as $n=\mathcal{F}(\mathcal{F}_{P}(c, o), \mathcal{F}_{R}(i, s), \mathcal{F}_{D}(v))$, where $\mathcal{F}$ denotes the integrating function. Note that, for convenience, we reuse some of these notations accented by a right arrow to represent the distributed representation of responding information or values. For example, we use $\vec{c}$ to indicate the distributed representation of $c$. Besides, we use $d$ to represent the dimension of vectors, and $W$, $b$, $U$ to represent trainable parameters.
In this paper, we consider symbolic music formatted as musical instrument digital interface~(MIDI) files, which consists of multiple chunks of information, directing the computing platforms to play the music properly. 
To accommodate the subsequent procedures, we conduct a series of modifications of the original MIDI files:
1) 
Inspired by ~\cite{hirai2019melody2vec,chuan2018modeling}, we normalize the keys of the song by transposing them to C Major.
This enables us to learn stable information about notes and music in the same tonal space. 
2) 
In theory, the melody track is made up of a sequence of single notes. 
As such, we define the track with the highest average pitch values as the melody track. We also split the accompaniment tracks into a series of sub-tracks by the octave to confine the chords to the same octave space. 
3) 
MIDI files encode music using three kinds of time units~(seconds, ticks, and beats). 
To avoid ambiguity, we normalize all kinds of time values into the music notation used in the scores, that is, the temporal value is represented in the form of semibreves (whole notes), minims ($1/2$ notes), crotchets ($1/4$ notes), etc.

\begin{figure}[t]
    \includegraphics[width=7.2cm]{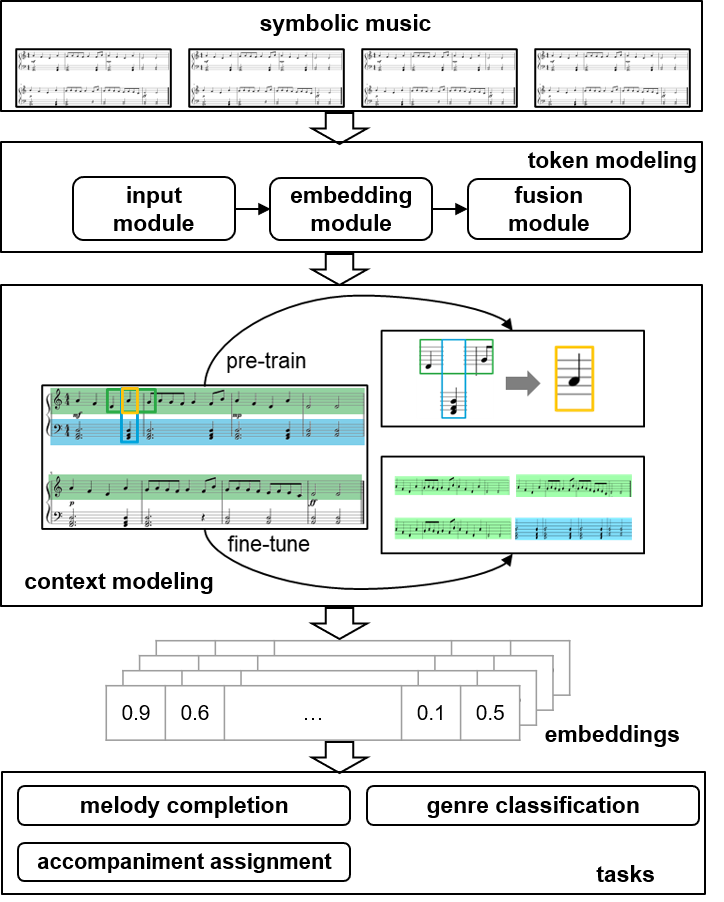}
    \caption{Architecture of PiRhDy}
    \label{fig:pipeline}
\end{figure}
\section{Methodology}
\label{sec:method}
We realize the proposed PiRhDy framework in two stages, as illustrated in Figure~\ref{fig:pipeline}. The first is a token modeling network to simultaneously fuse the pitch, rhythm, and dynamics information. 
This network consists of an input module extracting key features from note events based on a concise vocabulary, an embedding module encoding each feature into dense embeddings and a fusion module integrating all embeddings into a unified representation. 
The second is 
a context modeling network to capture the hidden relationship and distribution property of music at the sequence-level. 
In this network, the pretrained embeddings are smoothed over both horizontal (melodic) and vertical (harmonic) contexts. Apart from this, we introduce the fine-tune strategy on global contexts to inject long-term contextual information into the embeddings. 

\begin{figure}
    \includegraphics[width=6cm]{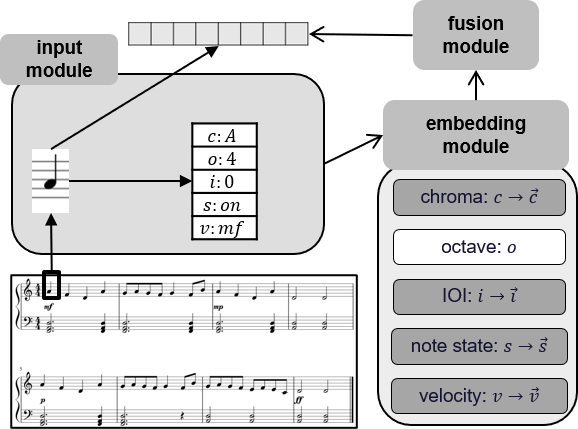}
    \caption{Token modeling network}
    \label{fig:token}
\end{figure}

\subsection{Token Modeling Network}
In this work, we treat the musical token as a single note event, which is the smallest meaningful variable of music. A note event is a combination of pitch, rhythm, and dynamics information. The basic idea of the token modeling network is to extract key features from a note event and encode such information into a unified embedding of this musical token, as displayed in Figure \ref{fig:token}.

\textbf{\textit{Input module}}
\label{sec:voc}

\begin{figure}[t]
        \includegraphics[width=8cm]{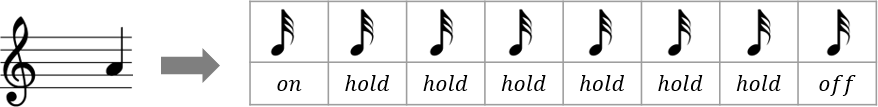}
        \caption{Example of note state}
        \label{fig:state}
    \end{figure}

The input module is to extract pitch, rhythm, and dynamics information from a music token. To this end, we define a comprehensive vocabulary consisting of all related information. Specifically, the entire vocabulary consists of three fundamental vocabularies. 
    \textbf{Chroma vocabulary} 
    is 
    defined as $\mathbb{C}=\{C,C\#,D,D\#,E, F, G,$ $G\#,A,A\#,B, R\}$,
    where $R$ denotes the chroma value of the rest note event~(\texttt{REST}). We combine chords into the chroma vocabulary because the simultaneous notes presented in a chord may all be represented in the chroma. For example, the vocabulary involves the C major chord $\{C, E, G\}$. 
    %
    \textbf{Velocity vocabulary} is 
    defined as $\mathbb{V} = \{pppp, ppp, pp, p, mp, mf,$ $f, ff, fff, ffff, R\}$,
    where the number of $p$ and $f$ indicate the level of softness and loudness respectively, $m$ indicates \textit{moderation} and $R$ represents the velocity of \texttt{REST}.
%
%
%
    \textbf{Note state vocabulary} is
     defined as $ \mathbb{S} =\{\text{\textit{on}}, \text{\textit{hold}}, \text{\textit{off}}, \text{\textit{r}}\}$,
    where \textit{on}, \textit{hold} and \textit{off} denote the start, sustain, and end actions
    of a note respectively; and $r$ represents the action of \texttt{REST}. As shown in Figure~\ref{fig:state}, if a note event is defined as a $1/32$ note long, the note duration of $A4$ is computed as $t_{\text{\textit{off}}}-t_{\text{\textit{on}}}$, note onset is $t_{\text{\textit{on}}}$, and IOI is $t'_{\text{\textit{on}}}-t_{\text{\textit{on}}}$, where $t'_{\text{\textit{on}}}$ represents the onset of the next note. Take the first $1/4$ note event in Figure~\ref{fig:token} as an example, the IOI value against itself is 0, against the second $1/4$ note event is $-1$, and so on. As such, we can get all rhythm-related information from the note state alphabet within only four unique symbols.

    As mentioned in section~\ref{sec:term}, the features of a single note event contain five dimensions including its chroma, octave, IOI, note state, and velocity. In practice, the IOI is computed from the current and a selected note's events (onsets). The chroma and octave values make up the pitch information, while the IOI, and note state features form the rhythm information. The velocity features constitute the dynamics information.

    \textbf{\textit{Embedding module}}

    The embedding module encodes all properties extracted from the input module into distributed vectors. For chroma, note state, and velocity, which are defined in the vocabulary, we use an embedding layer to encode the information into dense vectors~(i.e., $\vec{c}$, $\vec{s}$, and $\vec{v}$). The embedding layer outputs the product of the high-dimensional one-hot representations of chroma, velocity, and note state with an embedding matrix $W_E \in \mathbb{R}^{V\times d}$, where $V$ indicates the vocabulary size and $d$ is the dimension of vectors. Considering how the concept of IOI reflects the relative positions between node events, we use the trigonometric position encoding~\cite{vaswani2017attention} to obtain the IOI embeddings.
    Octave value~($o$) is not computed in this module and fed into the next module~(i.e., note module) directly.

    \textbf{\textit{Fusion module}}

\begin{figure}
    \centering
    \includegraphics[width=6cm]{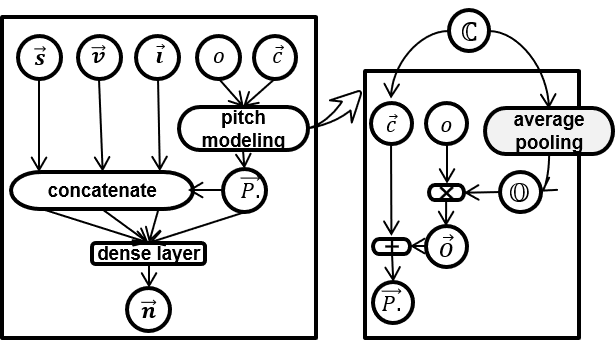}
    \caption{Fusion module}
    \label{fig:fusion}
\end{figure}

    The fusion module~(Figure~\ref{fig:fusion}) integrates vectors produced from the embedding module. Considering how pitch is composed of chroma and octave, we introduce a pitch modeling technique to obtain the dense representation of pitch~($\vec{P.}$) from $o$ and $\vec{c}$. We exploit the fact that musical notes that are linearly spaced on the MIDI scale are in fact logarithmically spaced on the frequency scale~\cite{bock2012polyphonic}. Assuming octave information is shared on all chromas, and map the average pooling on all chroma embeddings from $\mathbb{C}$ to get the octave representation $\mathbb{O}$. The octave representation~($\vec{o}$) of a note event is defined as $o\times\mathbb{O}$, where $\times$ is the scalar multiplication. Accordingly, the pitch representation is defined as $\vec{c}+\vec{o}$. We deploy a dense layer on the concatenation of $\vec{P.}$, $\vec{i}$, $\vec{v}$, and $\vec{s}$ to generate the unified representation of a note event.
\par

\subsection{Context Modeling Network}\label{sec:context}
The context modeling network smooths the embeddings on both melodic and harmonic contexts. To this end, we build a local context model and a global context model to capture short-term and long-term knowledge, respectively. In practice, we employ a hierarchical strategy which pretrains the embeddings on the local context model~(token-level), and fine-tunes these embeddings on the melodic and harmonic global contexts(sequence-level), separately. In this way, we can obtain the melody-preferred and harmony-preferred embeddings accordingly. 
\begin{figure}
    \includegraphics[width=8cm]{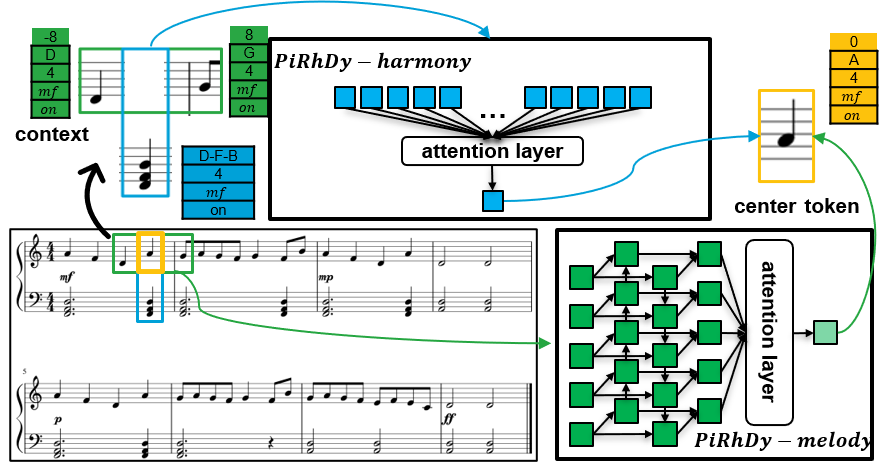}
    \caption{Local context module}
    \label{fig:local}
\end{figure}

\textbf{\textit{Local context module}}

Inspired by word2vec~\cite{mikolov2013distributed,mikolov2013efficient}, we develop a token-level module to handle local contextual information through sliding windows. The local context module~(Figure~\ref{fig:local}) aims to predict the center note event~(yellow part) from both melodic~(green part) and harmonic(blue part) contexts.

We define the state value for center note event from the melody track to be $on$. Melodic context is the set of note events with state values $on$ that surrounds the center note event from the melody track. Harmonic context comprise of note events with the same IOI value as the center note event from the accompaniment tracks. As such, the value of IOI is computed between the current note event and the center note event. 
Unlike the position value in word embeddings, the IOI in this module may not be continuous. Besides, it is not necessary to consider the constant IOI information when encoding the harmonic context. As illustrated in Figure~\ref{fig:local}, if we set a note event to be a $1/32$ note long, and take $\{0, A, 4, mf, \textit{on}\}$ as the center note event. The melodic context is $\{(-8, D, 4, mf, on), (8, G, 4, mf, on)\}$ and the harmonic context is $\{(D-F-B, 4, mf, on)\}$, both with window size set as 1. In practice, the harmonic window is always much larger than the melodic window.

The core of the module is a recurrent neural network (RNN)- and attention-based layers. The RNN-based layer
 captures the temporal information over the context and the attention layer is used to balance the importance of different note events.
We encode melodic context in a network~(denoted as~\textit{PiRhDy-melody}), where vectors from the token module are fed into $RnnL$ layers to get the temporal information of the melodic context. The output of $RnnL$ layers is followed by an $AttL$ layer to generate the hidden representation of the melodic context. Consider that the harmonic note events share the same IOIs, the network for harmonic context~(denoted as \textit{PiRhDy-harmony}) 
slightly modifies
the melodic network by removing the $RnnL$ layers.

\textbf{\textit{Global context module}}

The global context module is a sequence-level network smoothing embeddings on long-term contexts with the fine-tune strategy. Similar to the local context module, we define the center phrase as a phrase from the melody track, the melodic context as the following phrase of the center phrase, and the harmonic
context as the set of phrases from the accompaniment tracks corresponding to the center phrase. Either context is encoded into phrase representations through the \textit{PiRhDy-melody} network. Specifically, we model global melodic and harmonic contexts separately, so that we can learn the melody-preferred and harmony-preferred embeddings.

To encode global contexts, we design the period encoder and track encoder for melodic context and harmonic context, respectively. The period encoder starts from \textit{PiRhDy-melody}, followed by phrase-level $RnnL$s and  $AttL$ sequentially, to encode the melodic context into a dense representation of the period. This encoder is written as $\vec{per.} = AttL\{RnnL[\text{\textit{PiRhDy-melody}}({ph._k})_{k=1}^n]\}$, where $ph._k$ is a sequence of note event vectors for the $k^{th}$ phrase, $n$ is the number of input phrases, $\vec{per.}$ is the latent representation of the period. Following the local modeling network, the track encoder is a modified period encoder without the $RnnL$s. As such, the track encoder is formulated as $\vec{TRK} = AttL[\text{\textit{PiRhDy-melody}}({ph._k})_{k=1}^n]$, where $\vec{TRK}$ is the representation of vertical phrase-level context.

\subsection{{Optimization}}

    We aim to estimate the probability distribution of symbolic music from both melodic and harmonic contexts locally and globally. To this end, we formulate the loss function of local modeling as
\begin{equation}
    \small
    \mathcal{L} = f\Big(\sum\mathcal{L}_m, \sum\mathcal{L}_h\Big), 
    \label{eq:local}
\end{equation}  
\begin{equation}
    \small
    \mathcal{L}_m = -\sum_{-w_m\leq k\leq w_m, k\neq 0} \mathcal{P}\Big[(p_k, c_k, o_k, v_k, s_k)|(p_0, c_0, o_0, v_0, s_0)\Big],        
\end{equation}    
\begin{equation}
    \small
    \mathcal{L}_h = -\sum_{-w_h\leq k\leq w_h, k\neq 0} \mathcal{P}\Big[(c_k, o_k, v_k, s_k)|(c_0, o_0, v_0, s_0)\Big],    
\end{equation}
    where $\mathcal{L}_m$ denotes the loss calculated from the melodic contexts, $w_m$ is the sliding window size on melodic contexts, and $\mathcal{L}_h$ and $w_h$ serve the same functions as $\mathcal{L}_m$ and $w_m$ in the harmonic context, respectively.
    We introduce $f(*)$ as a weighting function to balance the importance of melodic and harmonic costs. As for global modeling, we aim to maximize the probability of $\sum\mathcal{P}(ph.*|ph.)$, where $ph.*$ is either the following phrase or an accompaniment phrase of a given phrase $ph.$, depending on the fine-tuning method.

    For fast and stable optimization, inspired by word2vec~\cite{mikolov2013distributed,mikolov2013efficient}, we adopt the negative sampling strategy for both pretraining and fine-tuning. Notably, towards local modeling, we consider a novel four-level method instead of sampling a negative sample for each positive sample. Practically, for every positive sample, we randomly replace one, two, three, and all features except for IOIs. In other words, there are four negative samples for a positive sample labeled as $[1,1,1,1]$. The optimization becomes a multi-label classification task than a binary classification task of word2vec. In this way, the model is sensitive to all the features during the training procedure.
    {
    As such, $\mathcal{L}$ can be estimated through the binary cross-entropy as
    \begin{equation}
        \label{eq:loss}
        \small
        \mathcal{L} = -\frac{1}{N}\sum\nolimits_{k=1}^{N}\sum\nolimits_{j=1}^{4}\Big[{y_k^j\log(\hat{y}_k^j) + (1 - y_k^j)\log(1 - \hat{y}_k^j)}\Big],
    \end{equation}
    where $N$ is the number of samples, $y$ is the ground truth value, and $\hat{y}$ is the predicted value.
}

\begin{table}
    \centering
    \caption{{Overview of the model names. LN indicates the line number. 
    }}
    \scalebox{0.75}{
    \begin{tabular}{|c|l|m{7cm}|}
        \toprule
        LN&\textbf{name} & \textbf{description}\\
        \midrule
        1&\textit{chroma}& the base model for the study of features only leverages the chroma features. Contexts involve both melody and harmony. The fusion operation follows the weighted function EQ.~\ref{eq:local}.\\
        \midrule
        2&\textit{chroma + octave}& leverages both chroma and octave features\\
        \midrule
        3&\textit{chroma + IOI}& leverages both chroma and IOI features\\
        \midrule
        4&\textit{chroma + note state}& leverage both chroma and note state features\\
        \midrule
        5&\textit{chroma + velocity}& leverages both chroma and velocity features\\
        \midrule
        \midrule
        6&\textit{PiRhDy-melody}&leverages all features on the melodic context\\
        \midrule
        7&\textit{PiRhDy-harmony}&leverages all features except for IOI on the harmonic context\\
        \midrule
        \midrule
        8&\textit{PiRhDy-AVG}&leverages all features on both melodic and harmonic contexts following the average operation, that is, reformulating EQ~\ref{eq:local} as $\mathcal{L}=\frac{1}{2}(\sum \mathcal{L}_h + \sum \mathcal{L}_v)$\\
        \midrule
        9&\textit{PiRhDy-WT}&leverages all features on both melodic and harmonic contexts following the weighted operation, that is, EQ~\ref{eq:local}\\
        \midrule
        \midrule
        10&\textit{PiRhDy-CP}& leverages cartesian product-based vocabulary \\
        \bottomrule
    \end{tabular}
    }
    \label{tab:model}
\end{table}
\begin{table}[t]
    \centering
    \caption{
    The effects of \emph{features} (LN 1-5), \emph{contexts} (LN 6-7), and \emph{fusion operations} (LN 8-9) demonstrated by binary cross-entropy values under different proportions of training data size (e.g., 20\% in the table means using 20\% of the training corpus). 
    }
    \vspace{-0.2em}
    \scalebox{0.8}{
    
            \begin{tabular}{|r|l|c|c|c|c|c|}
    \toprule
    \multicolumn{1}{|c|}{\multirow{2}{*}{LN}} & \multirow{2}{*}{model} & \multicolumn{5}{c|}{training data size} \\
\cmidrule{3-7}          & \multicolumn{1}{l|}{} & 20\%  & 40\%  & 60\%  & 80\%  & 100\% \\
    \midrule
    1     & \textit{chroma} & 0.5079 & 0.5078 & 0.5076 & 0.5076 & 0.5076 \\
    \midrule
    2     & \textit{chroma + octave} & 0.3639 & 0.3637 & 0.3636 & 0.3635 & 0.3636 \\
    \midrule
    3     & \textit{chroma + IOI} & 0.5078 & 0.5074 & 0.5073 & 0.5069 & 0.5071 \\
    \midrule
    4     & \textit{chroma + note state} & 0.3445 & 0.344 & 0.3439 & 0.3436 & 0.3426 \\
    \midrule
    5     & \textit{chroma + velocity} & 0.4049 & 0.4045 & 0.4049 & 0.4048 & 0.4048 \\
    \midrule
    \midrule
    6     & \textit{PiRhDy-melody} & \textbf{0.0645} & \textbf{0.0627} & \textbf{0.0627} & \textbf{0.0619} & 0.0619 \\
    \midrule
    7     & \textit{PiRhDy-harmony} & 0.3693 & 0.3534 & 0.3617 & 0.3613 & 0.3548 \\
    \midrule
    \midrule
    8     & \textit{PiRhDy-AVG} & 0.1384 & 0.1364 & 0.1359 & 0.1357 & 0.1362 \\
    \midrule
    9     & \textit{PiRhDy-WT} & 0.0668 & 0.0633 & \textbf{0.0627} & \textbf{0.0619} & \textbf{0.0617} \\
    \bottomrule
    \end{tabular}%

    }

    \label{tab:feature}%
  \end{table}%

\section{Settings}
Following melody2vec~\cite{hirai2019melody2vec}, we use the Lakh MIDI dataset~\cite{raffel2016learning}, which is a collection of 178,561 unique MIDI files. For efficient learning, we omitted melodic phrases with less than 75\% of valid note events, harmonic phrases with less than 50\% of valid note events, and songs with less than two periods. After preprocessing, we obtained 643 unique chords. Thus, the sizes of chroma, velocity and note state vocabularies are 656~(13+643), 11 and 4, respectively, resulting in an entire vocabulary containing 671 symbols, which is much lesser than melody2vec~(286,003 symbols).

\section{Experiment \uppercase\expandafter{\romannumeral1}: Study of PiRhDy}
\label{sec:study}
To comprehensively examine the embeddings, we designed two sets of empirical evaluations. In this section, we conduct a detailed study of each strategy set in the local context module by decomposing the proposed framework. In the next section (i.e., Section~\ref{sec:down}), we further examine the robustness and effectiveness of our embeddings fine-tuned on global contexts by designing both sequence-level and song-level downstream tasks.
To analyze the impact of strategies proposed in this work, we randomly split the corpus into training~(90\%) and testing~(10\%), and design a detailed study about features,
contexts,
fusion operations,
training data size,
and vocabularies. {All these studies are built on the local context module~(section~\ref{sec:context}). Table~\ref{tab:model} describes the model names used in this section. 
Following \cite{madjiheurem2016chord2vec}, we report the binary cross-entropy values of EQ.~(\ref{eq:loss}) on the test data
in Tables~\ref{tab:feature} and \ref{tab:embedding}. 
}

\textbf{\textit{Study of features: }}
As explained in section~\ref{sec:term}, a single note event consists of chroma, octave, IOI, note state, and velocity features. We deploy a base model~(denoted as \textit{chroma}), which only leverages the chroma feature input. We choose chroma as the base feature because it is the only feature used by all related works. The performance of other features are studied by adding them to base model.
For instance, we use the model~(\textit{chroma + octave}) to denote the adding of octave values to chroma as input. Note that this note representation is equal to the pitch representation. 
{
The remaining features are studied in the same way as the octave features. Apart from this, \textit{PiRhDy-WT} is the corresponding model that leverages all features~(i.e., chroma, octave, IOI, note state, and velocity).
}


{From the top part~(LN 1-5) of Table~\ref{tab:feature}, we can see that the addition of note state to the chroma model achieves the best performance, even though it contains only four symbols. The addition of octave gives the next best results, indicating the significance of pitch information~(the combination of chroma and octave) in music. The addition of velocity, expresses the loudness of each note provides the next best improvement. The least improvement is seen in the addition of IOI, which expresses the relative position of note events from the melodic context. Both velocity and IOI slightly contribute to predicting the center note event. We conjecture that this is because 
the dataset contains little variation in velocity changes,
 limiting our learning of its variation. 
 IOI values were insufficient to capture a significant pattern. Further to this, from the last line~(\textit{PiRhDy-WT}) of Table~\ref{tab:feature}, we can see that even better results, indicating that it is best to use all features together.}


\textit{Observation 1:} It is beneficial to integrate pitch, rhythm, and dynamics information, namely, fusing chroma, octave, IOI, note state, and velocity features.

\textbf{\textit{Study of contexts: }}
{There are two types of contexts, namely, melodic context from the melody track and harmonic context from the accompaniment track. Towards the study of these contexts, we leverage all features in the melodic context. Note that, however, only the chroma, octave, note state, and velocity features are leveraged in the harmonic context. This is because the IOI feature is the same for all note events of the harmonic context.} 

From the middle part~(LN 6-7) of Table~\ref{tab:feature}, we can see that \textit{PiRhDy-melody} always works better than \textit{PiRhDy-harmony}. This indicates that the center note event is more related to melodically surrounding note events than harmonically surrounding ones.


\textit{Observation 2:} Melodic context contributes more to the center note event than the harmonic context. Besides, these contexts may possess different weights for the local distribution of music.

\textbf{\textit{Study of fusion operations: }} 
Based on \textit{Observation 2}, we defined a weighted fusion operation in EQ.~(\ref{eq:local}). To study the impact of EQ.~(\ref{eq:local}), we conduct the average operation on the outputs of \textit{PiRhDy-melody} with \textit{PiRhDy-harmony} as a baseline method,
denoted as \textit{PiRhDy-AVG}. Similarly, we refer to the model following the weighting function EQ.~(\ref{eq:local}) as \textit{PiRhDy-WT}. The results are presented in the bottom part~(LN 8-9) of Table~\ref{tab:feature}.

\textit{PiRhDy-WT} works better than the average operation to fuse the results encoded from melody and harmony. This implies that melodic and harmonic contexts have different importances when predicting the center note event, and confirms \textit{Observations 2} again. 
The performances of \textit{PiRhDy-melody} is inferior to \textit{PiRhDy-WT}, but superior to \textit{PiRhDy-AVG}. This suggests it is necessary to find an efficient way to integrate melodic and harmonic knowledge.

\textit{Observation 3:} Melodic and harmonic contexts contribute differently to predicting the center note event. It is necessary to balance the importance of melodic and harmonic knowledge appropriately.

\textbf{\textit{Study of training data size: }}
We conduct the study of training data size~(Table \ref{tab:feature}) on all feature-~(e.g., \textit{chroma}), context-~(e.g., \textit{PiRhDy-melody}) and fusion operation-related~(e.g., \textit{PiRhDy-AVG}) models
by gradually increasing the percentage of training corpus from 20\% to 100\%, while the testing data size remains the same.

Generally, a larger training data size produces better performances, and feature-related models are converge more readily than others. For example, \textit{chroma} converges with 60\% of the training corpus, while \textit{PiRhDy-melody} needs more than 80\% of the corpus to converge. We conjecture that this is because the learned information of feature-related models is less than that of others. Note that, when no more than 40\% of the training data is used, \textit{PiRhDy-melody} is the best approach; otherwise, \textit{PiRhDy-WT} is the best. We argue this is because significant amounts of harmonic information is needed for sufficient coverage in training. Consequentially, inefficient harmonic information decreases the performances of approaches based on both melodic and harmonic contexts.

\textit{Observation 4:} Larger corpus can produce more efficient embeddings, and the improvements shrink with larger data size.

\textbf{\textit{Study of vocabularies:}}{
    As explained in section~\ref{sec:voc}, we build our feature-based vocabulary as a combination of chroma, velocity, and note state vocabularies. In this way, a music token~(note event) is represented as a five-dimensional variable, involving chroma, octave, IOI, note state, and velocity features. We learn the embeddings of features and obtain the token representation by combining such embeddings accordngly. A natural alternative is to build the vocabulary on the cartesian products of chroma, octave, note state and velocity features, and embed such cartesian products into vectors. Similar approaches have been explored in \cite{herremans2017modeling, chuan2020context}, yet only using the pitch feature. We name the model using cartesian product-based vocabulary \textit{PiRhDy-CP}, and re-conduct the \textit{PiRhDy-WT} model on 1\%, 5\%, and 10\% of the training corpus. We compare the effectiveness of embeddings derived from the feature-based vocabulary and that derived from the cartesian product-based vocabulary in Table~\ref{tab:embedding}. Note that, in theory, the vocabulary size of \textit{PiRhDy-CP} is 346,368~(656 chromas, 12 octaves, 11 velocities, and 4 note states), of which we find 216,031 unique occurrences in the entire corpus.
}

\textit{PiRhDy-WT} works better than \textit{PiRhDy-CP} in all conditions, especially when training data size is small. In detail, \textit{PiRhDy-WT} can produce remarkable 
and almost converging 
results when the percentage is more than 20\%. However, the seemingly competitive results made by \textit{PiRhDy-CP} need more than 80\% of training data. In addition, the parameters of \textit{PiRhDy-CP}~(56,723,500) is more than 30 times larger than \textit{PiRhDy-WT}~(1,854,252). This means that the former is more difficult to train than the latter.

\textit{Observation 5:} The feature-based vocabulary is better than the cartesian product-based one on both robustness and effectiveness.

\begin{table}[t]
  \centering
  \caption{Study of vocabularies}
  \scalebox{0.75}{
   \begin{tabular}{|l|c|c|c|c|c|c|c|c|}
    \toprule
    datasize & 1\%   & 5\%   & 10\%  & 20\%  & 40\%  & 60\%  & 80\%  & 100\% \\
    \midrule
    \textit{PiRhDy-CP} & 0.4536 & 0.4106 & 0.2805 & 0.2004 & 0.1353 & 0.1172 & 0.0959 & 0.0804 \\
    \midrule
    \textit{PiRhDy-WT} & 0.3395 & 0.1512 & 0.1017 & 0.0668 & 0.0633 & 0.0627 & 0.0619 & 0.0617 \\
    \bottomrule
    \end{tabular}%
  }
  \label{tab:embedding}%
\end{table}%


\section{Experiment \uppercase\expandafter{\romannumeral2}: Downstream Tasks}
\label{sec:down}


% ...
In section~\ref{sec:study}, we studied the strategies set in the local context module. In this section, we deploy sequence- and song-level downstream tasks to evaluate the embeddings fine-tuned in the global context, namely, the PiRhDy embeddings. We use \textit{PiRhDy-WT}, the best performing local modeling strategy as discussed in section~\ref{sec:study}.
To clarify, we denote the embeddings fine-tuned on global melodic context as \textit{PiRhDy\_GM}~(i.e., melody-preferred embeddings), and embeddings fine-tuned on harmonic context as \textit{PiRhDy\_GH} (i.e., harmony-preferred embeddings). We evaluate these embeddings on the melody completion, accompaniment assignment and genre classification tasks. Both \textit{PiRhDy\_GM} and \textit{PiRhDy\_GH} are pretrained embeddings and directly fed into task-specific classifiers in a plug-and-play manner without further processing.
\subsection{Baseline Models}
According to our survey, all existing contextual embeddings~\cite{huang2016chordripple,madjiheurem2016chord2vec,herremans2017modeling,chuan2020context,alvarez2019distributed,hirai2019melody2vec} for symbolic music are based on word2vec~\cite{mikolov2013efficient,mikolov2013distributed}, and vary in the surface formats. For an efficient comparison, we choose melody2vec~\cite{hirai2019melody2vec} as our baseline model, because it is the state-of-the-art technique, leveraging both pitch and duration information, while also utilizing the Lakh dataset~\cite{raffel2016learning} as its corpus. 
Together with the matrix-based approaches, we summarize the baseline models as
\begin{itemize}[leftmargin=*]
    \item \textbf{melody2vec\_F}~\cite{hirai2019melody2vec}, melody2vec embeddings with the forward maximum matching algorithm for motif segmentation.
    \item \textbf{melody2vec\_B}~\cite{hirai2019melody2vec}, melody2vec embeddings with the backward maximum matching algorithm for motif segmentation.
    \item \textbf{tonnetz}~\cite{chuan2018modeling}, the music bar is represented as a tonnetz matrix.
    \item \textbf{pianoroll}~\cite{dong2018pypianoroll}, the music bar is represented as a time-pitch matrix.
\end{itemize}
For melody2vec, we leverage the pretrained embeddings released by \cite{hirai2019melody2vec}. For tonnetz and pianoroll, we apply convolutional neural networks
following \cite{chuan2018modeling} and \cite{dong2018pypianoroll}, respectively. 
\subsection{Melody Completion}
\begin{figure}
    \includegraphics[width=8cm]{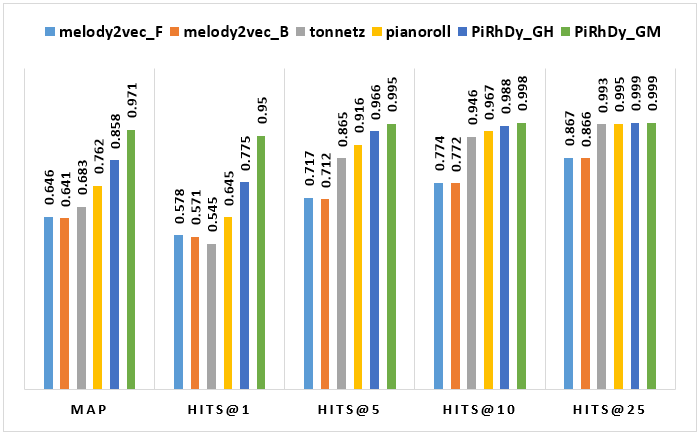}
    \caption{Results of melody completion task}
    \label{fig:next}
\end{figure}

Inspired by \cite{malmi2016dopelearning}, 
we design a melody completion task defined as follows: ``\textit{Given a melodic phrase ($ph._q$) and a set of candidate melodic phrases ($[ph.]_{melody}$), find its most related consecutive phrase from $[ph.]_{melody}$.}''
This  
is a sequence-level~(period-level) evaluation task, with a concrete application of \textit{PiRhDy-melody}. There are 1,784,844 pairs in the training dataset, and 199,270 pairs in the testing dataset. For each $ph._q$ in the training dataset, we randomly select a phrase
to form a negative sample.
For the testing dataset, we generate 49 negative samples for each positive sample. 
We use mean average precision~(MAP) and hits@k~(k=1, 5, 10, 25, indicating the rate of the candidates containing the correct next phrase) as evaluation metrics, and report the results in Figure~\ref{fig:next}.\par
Our embeddings~(\textit{PiRhDy\_GH }and \textit{PiRhDy\_GM}) perform better than melody2vec and tonnetz, suggesting that PiRhDy can capture melodic continuity. It is worth mentioning that \textit{PiRhDy\_GH} is outperformed by \textit{PiRhDy\_GM}, indicating that the vertical distribution is not as significant as the horizontal distribution of music. Thus, it is necessary to model globally vertical and harmonic contexts separately in specific tasks. 
\begin{figure}
    \includegraphics[width=6cm]{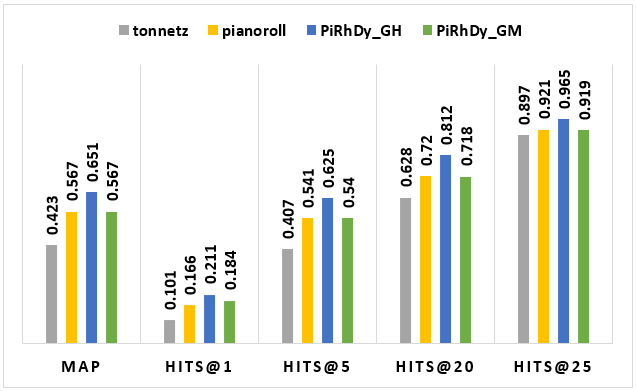}
    \caption{Results of accompaniment assignment task}
    \label{fig:acc}
\end{figure}
\subsection{Accompaniment Assignment}
Similar to the melody completion task, we formulate the accompaniment assignment task as follows: ``\textit{Given a melodic phrase ($ph._q$) and a set of candidate harmonic phrases ($[ph.]_{harmony}$), find its most related accompaniment phrase from $[ph.]_{harmony}$.}''
This
is also a sequence-level~(track-level) task, with a concrete application of \textit{PiRhDy-harmony}. There are 7,890,554 phrase pairs in the training set, and 200,300 samples in the testing set. For each $ph._q$, there are more than one correct candidates from $[ph.]_{harmony}$. Thus, in the training dataset, we randomly generate a negative sample for each positive sample. In the testing set, for each $ph._q$ with $N$ correct accompaniment phrases, we randomly generate $(50-N)$ phrases as negative samples.
We use MAP and hits@k as metrics, and report the results in Figure~\ref{fig:acc}. We omit the melody2vec approach~\cite{hirai2019melody2vec} because its embeddings only work for melodic phrases.\par
We observe that both \textit{PiRhDy\_GH }and \textit{PiRhDy\_GM}, 
perform better than baseline models. This demonstrates that \textit\textit{PiRhDy-WT} is best at capturing vertical continuity of music. We also observe that even \textit{PiRhDy\_GM}, which only learns vertical information from the local context, gains remarkable results as compared to the results of \textit{PiRhDy\_GH} in the next phrase prediction task. This implies that the relationship between locally and globally vertical context is stronger than that of the horizontal context.

\subsection{Genre Classification}
\begin{table}[t]
    \centering
    \caption{Results of genre classification task}
    \scalebox{0.8}{
      \begin{tabular}{|l|c|c|c|c|}
      \toprule
      \multirow{2}{*}{Model} & \multicolumn{2}{c|}{Top-MAGD} & \multicolumn{2}{c|}{MASD} \\
  \cmidrule{2-5}    \multicolumn{1}{|c|}{} & \multicolumn{1}{c|}{AUC-ROC} & \multicolumn{1}{c|}{F1} & \multicolumn{1}{c|}{AUC-ROC} & \multicolumn{1}{c|}{F1} \\
      \midrule
      SIA  & 0.753 & 0.637 & 0.761 & 0.455 \\
      \midrule
      P2   & 0.816 & 0.649 & 0.815 & 0.431 \\
      \midrule
      melody2vec\_F & 0.885 & 0.649 & 0.777 & 0.299 \\
      \midrule
      melody2vec\_B & 0.883 & 0.647 & 0.776 & 0.293 \\
      \midrule
      tonnetz & 0.871 & 0.627 & 0.794 & 0.253 \\
      \midrule
      pianoroll & 0.876 & 0.64  & 0.774 & 0.365 \\
      \midrule
     \textit{PiRhDy\_GH} & 0.891 & 0.663 & 0.782 & 0.448 \\
      \midrule
     \textit{PiRhDy\_GM} & \textbf{0.895} & \textbf{0.668} & \textbf{0.832} & \textbf{0.471} \\
      \bottomrule
      \end{tabular}%
    }
    \label{tab:genre}%
  \end{table}%
  
Genre classification is a song-level task. Following~\citeauthor{10.1145/3273024.3273035}~\cite{10.1145/3273024.3273035}, we conduct evaluations on TOP-MAGD and MASD datasets~\cite{Schindler2012Facilitating}. 
TOP-MAGD contains 22,535 files of 13 genres while MASD contains 17,785 files of 25 styles. We use AUC-ROC and F1 scores as metrics, and 5-fold cross-validation for evaluation in line with~\cite{10.1145/3273024.3273035,DBLP:conf/ismir/OramasNBS17}. Table~\ref{tab:genre} presents the results of different methods. SIA and P2 are baseline models proposed in \cite{10.1145/3273024.3273035}.\par
As can be seen from Table~\ref{tab:genre},
\textit{PiRhDy\_GM}
outperforms all methods on both metrics and both datasets, while  \textit{PiRhDy\_GH} comes in second.
This indicates that our proposed embeddings are also suitable for song-level tasks in addition to period- and track-level tasks. The more significant performances of \textit{PiRhDy\_GM} against \textit{PiRhDy\_GH} suggests that melody outweighs harmony in music genre. The results on MASD are less remarkable than on TOP-MAGD, because the styles are sub-genres of genres in TOP-MAGD. In other words, the genre classification on MASD is more challenging. Both \textit{PiRhDy\_GH }and \textit{PiRhDy\_GM} generate competitive results,  showing the scalability and robustness of PiRhDy embeddings. 

\section{Conclusion}
In this paper, 
we proposed PiRhDy as a comprehensive approach to embed music. This approach is built on a hierarchical strategy, understanding music from token-level to sequence-level. We designed a token modeling network to fuse key features into unified token representations and a context modeling network~(sequence-level) to smooth embeddings across global contexts. The experimental results consistently validated the robustness and effectiveness of our pretrained PiRhDy embeddings. 

To the best of our knowledge, PiRhDy is the first approach 
that learns music embeddings of 
rich musical features and 
draws knowledge from both melodic and harmonic contexts. 
On the strength of the encouraging performance, we confidently believe that the PiRhDy embeddings will advance future developments of computational musicology at symbolic level. 
As a plug-and-play tool, PiRhDy embedddings can be easily applied to any neural music processing application where symbolic music modeling is needed.
We plan to further explore embeddings smoothed on rhythm and tonal patterns, which is analogous to syntax in natural languages. This study can help to better understand and interpret the structural characteristics of music. Another future direction is to inject audio content into the learning process to further enable embeddings smoothed on multimodal content. 
Additionally, we will further explore the potential of PiRhDy in a broader range of tasks such as music similarity matching and music generation.
\section{Acknowledgments}
This work was supported in part by the Ministry of Education Humanities and Social Science project, China under grant 16YJC790123 and National Research Foundation, Singapore under its International Research Centres in Singapore Funding Initiative. Any opinions, findings and conclusions or recommendations expressed in this material are those of the author(s) and do not reflect the views of National Research Foundation, Singapore.
\bibliographystyle{content/ACM-Reference-Format}
\bibliography{content/ref}

\end{document}